\documentclass[
  aps,
  pre,
  reprint,
  twocolumn,
  superscriptaddress,
]{revtex4-2}

\usepackage{graphicx}
\usepackage{amsmath}
\usepackage[textsize=tiny]{todonotes}
\usepackage{siunitx}
\usepackage{tikz}
\usepackage[dvipsnames]{xcolor}
\usepackage{hyperref}

\DeclareRobustCommand{\mean}[1]{\ensuremath{ { \langle #1 \rangle} }}
\DeclareRobustCommand{\vect}[1]{\ensuremath{\boldsymbol{#1}}}
\newcommand{\dd}{\ensuremath{\,\mathrm{d}}} % Differential d

\DeclareRobustCommand{\revision}[1]{{{#1}}}

\begin{document}

% Use the \preprint command to place your local institutional report
% number in the upper righthand corner of the title page in preprint mode.
% Multiple \preprint commands are allowed.
% Use the 'preprintnumbers' class option to override journal defaults
% to display numbers if necessary
%\preprint{}

%Title of paper
\title{Diffusion in the Inverted Triangular Soft Lorentz Gas}

% repeat the \author .. \affiliation  etc. as needed
% \email, \thanks, \homepage, \altaffiliation all apply to the current
% author. Explanatory text should go in the []'s, actual e-mail
% address or url should go in the {}'s for \email and \homepage.
% Please use the appropriate macro foreach each type of information

% \affiliation command applies to all authors since the last
% \affiliation command. The \affiliation command should follow the
% other information
% \affiliation can be followed by \email, \homepage, \thanks as well.

\author{Esko Toivonen}
\email{esko.toivonen@tuni.fi}
\affiliation{Computational Physics Laboratory, Tampere University, P.O. Box 600, FI-33014 Tampere, Finland}

\author{Aleksi Majaniemi}
\affiliation{Computational Physics Laboratory, Tampere University, P.O. Box 600, FI-33014 Tampere, Finland}

\author{Rainer Klages}
\affiliation{Centre for Complex Systems, School of Mathematical Sciences, Queen Mary University of London, Mile End Road, London E1 4NS, United Kingdom}
\affiliation{London Mathematical Laboratory, 8 Margravine Gardens, London W6 8RH, UK}

\author{Esa Räsänen}
\email{esa.rasanen@tuni.fi}
\affiliation{Computational Physics Laboratory, Tampere University, P.O. Box 600, FI-33014 Tampere, Finland}

\date{\today}

\begin{abstract}
We investigate diffusion in a two-dimensional inverted soft Lorentz gas, where attractive Fermi-type potential wells are arranged in a triangular lattice. This configuration contrasts with earlier studies of soft Lorentz gases involving repulsive scatterers. By systematically varying the gap width and softness of the potential, we explore a rich landscape of diffusive behaviors. We present numerical simulations of the mean squared displacement and compute diffusion coefficients, identifying tongue-like structures in parameter space associated with quasiballistic transport. Furthermore, we develop an extension to the Machta-Zwanzig approximation that incorporates correlated multi-hop trajectories and correct for the influence of localized periodic orbits. Our findings highlight the qualitative and quantitative differences between inverted and repulsive soft Lorentz gases and offer new insights into transport phenomena in smooth periodic potentials.
\end{abstract}

\maketitle

\section{Introduction}

Over the past two decades, two-dimensional (2D) materials have revolutionized condensed matter physics by enabling the exploration of quantum phenomena in systems with reduced dimensionality~\cite{2D}. In small-scale electronic systems --where only a few degrees of freedom play a role -- transport often occurs far from equilibrium and exhibits nonlinear, sometimes chaotic, behavior. This gives rise to striking effects such as branched electron flow~\cite{Heller2021,Daza2021, graf2024chaos} and anomalous diffusion~\cite{KRS08,MJJCB14,ZDK15,Oliveira2019}, where particle motion deviates from classical expectations. 

To explore these transport phenomena, the Lorentz gas~\cite{Sza00,Klages_book,Dettmann_2014} is often employed as a model to describe a single particle moving through a fixed array of scatterers. This simple framework has become fundamental for studying complex dynamical behavior, including energy dissipation, sensitivity to initial conditions, and manifestations of chaos~\cite{Do99,Gaspard1998}. A more realistic variant -- the “soft” Lorentz gas -- replaces the hard-wall scatterers with smooth, Fermi-type potential profiles arranged in various geometries, such as triangular~\cite{PRL} or square~\cite{toivonen2025} lattices. This version is particularly well-suited for modeling two-dimensional electronic systems like artificial graphene~\cite{Gomes2012,Rasanen2012,Paavilainen2016,Polini2013}, where the underlying potential landscape is continuous. These systems exhibit both normal and anomalous diffusion, with extreme sensitivity to the model parameters~\cite{PRL,toivonen2025}.

In this work, we investigate the {\em inverted} soft Lorentz gas, in which the repulsive potential bumps are replaced by attractive potential wells -- similar to the experimental setup described in Ref.~\cite{experiment}. \revision{Moreover, triangular optical lattices with tunable potential landscapes have already been realized with ultracold atoms~\cite{Becker2010}, further underlining the potential experimental relevance of our model.} This modification leads to markedly different system behavior compared to the conventional square or triangular lattices with repulsive scatterers. In particular, our inverted setup exhibits an infinite horizon for all parameter regimes where particles possess enough energy to escape the wells and travel to infinity~\cite{machta_zwanzig,klages2000,dettmann2012}. In soft-potential systems, variations in the potential landscape cause speed fluctuations along trajectories, leading to complex dynamics and regular structures in phase space~\cite{GZR88,turaev1998islands,RKT99}. 
An infinite horizon complicates analysis further; in hard-wall \revision{two-dimensional} systems, it rules out normal diffusion~\cite{Sza00,Klages_book,Dettmann_2014}\revision{, while
in higher dimensions the situation is more complicated~\cite{dettmann2012}. In} soft systems such horizons similarly challenge conventional diffusion approximations. In particular, the Machta–Zwanzig (MZ) approximation, which assumes uncorrelated hops between unit cells~\cite{machta_zwanzig}, breaks down in infinite-horizon lattices where normal diffusion is ill-defined\revision{~\cite{toivonen2025}.
The main assumption behind the MZ approximation is that hops between traps are uncorrelated. In infinite horizon lattices, however, channels appear leading to particle movement along straight lines across (infinitely) many traps. This prevents randomisation in each trap, which leads to non-Markovian correlations in the dynamics. 
}

Building on previous work by some of the present authors~\cite{toivonen2025}, we employ an approximation that explicitly incorporates correlations between transitions across unit cells. This approach is benchmarked against both the conventional MZ approximation and numerical computations of the diffusion coefficient, with additional corrections introduced to account for the influence of confined orbits (COs)~\footnote{This term is not related to carbon monoxide.} which are confined in a single potential well. An important subset of these orbits are the localized periodic orbits (LPOs), which are regular in addition to confinement. Our results reveal qualitative similarities with soft Lorentz gases on triangular and square lattices, including the emergence of both normal and anomalous diffusion and a pronounced sensitivity to model parameters. Nonetheless, the specific form of the potential introduces key differences that significantly impact the accuracy of various approximation methods.

This paper is organized as follows. In Sec.~\ref{sec:model}, we introduce the numerical scheme utilized in the simulations. In Sec.~\ref{sec:theory_pred}, we discuss the diffusive properties of the system as a function of system parameters and introduce the approximations we use.
We present our results in Sec.~\ref{sec:results}, and conclude the article in Sec.~\ref{sec:conclusion}.

\section{Model and methodology}\label{sec:model}

Our inverted soft Lorentz gas is defined as a triangular arrangement of circular potential wells (see Fig.~\ref{fig:2dpot}), modeled with a smooth Fermi potential
\begin{equation}
    V_\mathrm{s}(\boldsymbol{r})=-\frac{V_0}{1+\exp\left(\frac{\mid\boldsymbol{r}\mid-r_0}{\sigma}\right)},
\end{equation}
where~$r_0$ is the effective radius of each potential well and~$\sigma$ is a softening parameter.
In the lattice, the total potential affecting a particle is calculated as
\begin{equation}\label{eq:potential}
    V(\boldsymbol{r})= \revision{V_0}+\sum_{n}V_\mathrm{s}(\boldsymbol{r}-\boldsymbol{r}_n),
\end{equation}
where~$\boldsymbol{r}_n$ is the location of the center of the $n$th potential well and the "+$V_0$" ensures that the potential is above zero in parameter space that we are interested in.
For simplicity, we choose~$V_0 = 1$.
In Hartree atomic units ($r_0 = m = 1$), we set the total energy of the particle to $E=1/2$.
As the potential height is at most one, we expect this energy choice to be representative of the dynamics in parameter space where the particles can cross between potential wells but cannot move entirely freely in the space.
\revision{In the present work, we therefore varied only the gap width $w$ and the softness parameter $\sigma$ to access different dynamical regimes. We note, however, that varying the particle energy can also modify these regimes; see Ref.~\cite{GilGallegos2019} for a detailed study.} Furthermore, we define the unit cell side length as~$L$, given by~$L = 2r_0 + w$, where~$w$ is the gap width between potential wells.

We study the 2D diffusion coefficient~$D$, given by
\begin{equation}\label{eq:dcoef}
    D = \lim_{t\to\infty} \frac{\langle(\boldsymbol{r}(t) - \boldsymbol{r}(0))^2\rangle}{4t}
\end{equation}
with the angle brackets denoting a configurational ensemble average, forming the mean squared displacement (MSD) as~$\langle(\boldsymbol{r}(t) - \boldsymbol{r}(0))^2\rangle$.
Here,~$\boldsymbol{r}(t)$ refers to the position of a particle at time~$t$.
We note that Eq.~\eqref{eq:dcoef} is only valid for \emph{normal} diffusion where the MSD grows linearly in time.
Anomalous diffusion -- superdiffusion and subdiffusion -- is obtained when MSD~$\propto t^\alpha$ with $\alpha > 1$ and $\alpha < 1$, respectively.
\revision{
The main sources of anomalous diffusion are COs, LPOs and quasiballistic orbits.
We define quasiballistic orbits as regular trajectories that propagate infinitely toward one direction. However, we note that they may display periodic small-scale features along the way (see Fig.~\ref{fig:ballistic} below).}

For computing the~$D$ with Eq.~\eqref{eq:dcoef}, we require a sufficiently large ensemble of simulations.
For this purpose, we employ the Bill2D software package~\cite{bill2d}.
We utilize the sixth-order symplectic integrator~\cite{theintegrator} with a timestep of~$\Delta t = 10^{-3}$, as this combination has been shown to produce the most accurate conservation of energy within the parameter space in our simulations.~\cite{toivonen2025}
To accurately compute the potential values affecting a particle, we form a 3-by-3 unit cell grid and place the particle in the centermost unit cell.
Unless otherwise specified, \revision{as the invariant density of the system is not known,} we randomize the initial positions and propagation angles of the particles uniformly according to the energetically allowed phase space determined by the constant energy~$E=1/2$.

\section{Theoretical predictions}\label{sec:theory_pred}

We begin by exploring the parameter space to identify the regions where diffusion is expected to occur, followed by the introduction of analytical approximations for the diffusion coefficient $D$.

\subsection{Parameter space}

\begin{figure}[htb]
    \centering
    \includegraphics{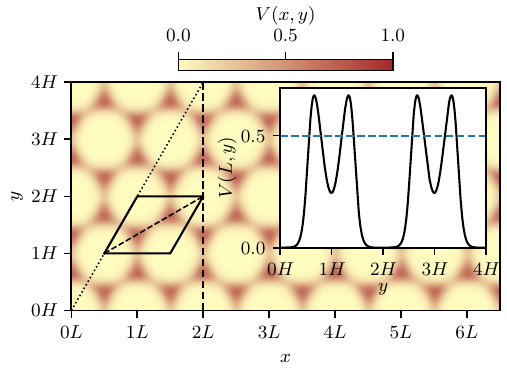}
    \caption{
        Two-dimensional representation of the potential wells of the inverted triangular soft Lorentz gas.
        The parallelepiped represents the unit cell employed in the simulations, and the dashed line inside this unit cell shows the Poincaré surface of section (PSOS).
        The dotted line indicates an example of an infinite horizon in the system.
        The dashed vertical line represents the cross-section depicted in the inset.
        In the inset, the dashed horizontal blue line represents the constant energy,~$E=1/2$.
        $L$ represents the length of the side of the unit cell and $H$ its height.
    }
    \label{fig:2dpot}
\end{figure}

Due to the inverted nature of the potential, which is now characterized by attractive wells rather than repulsive bumps, the relevant regions of parameter space differ notably from those encountered in conventional Lorentz gas setups. This distinction is illustrated in Fig.~\ref{fig:2dpot}. 
We can deduce that as a function of the gap width~$w$ and system softness~$\sigma$, several regions of differing behavior appear.
First, when the potentials are far away from each other (large~$w$) and/or the potential walls are hard (small~$\sigma$), particles become confined within individual wells. In this limit, the system effectively behaves as a set of isolated soft circular billiards \revision{slightly perturbed by the surrounding wells, and exhibits no diffusion.}
While we are not aware of direct studies of this specific configuration, Ref.~\cite{kroetz2016} notes that soft elliptical billiards with low eccentricity lack a stochastic layer, which suggests the absence of chaotic dynamics.
However, while not utilizing the exact Fermi potential, Ref.~\cite{gonzales_arxiv} suggests that even billiards with soft walls might exhibit chaos.

\begin{figure}[htb]
    \centering
    \includegraphics{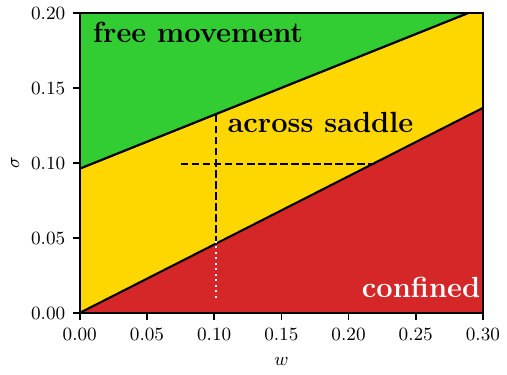}
    \caption{
        Parameter space highlighting regions of different dynamical behavior.
        The vertical cross sections marked with dotted white and dashed black lines denote further analysis shown in Figs.~\ref{fig:dcoef_s}(a) and (b), respectively.
        The horizontal cross section marked with a dashed line denotes analysis presented in Fig.~\ref{fig:dcoef_w}.
        Any similarities to the flag of the Republic of the Congo are coincidental.
    }
    \label{fig:paramspace}
\end{figure}

As the wells get closer together or the system becomes softer, the particles can eventually move between wells through the saddle points, which is seen in the cross-section shown in the inset of Fig.~\ref{fig:2dpot}.
We can obtain an approximative equation for this limit by considering two adjacent potential wells. We obtain a condition
\begin{equation}
    \sigma \gtrsim \frac{1}{2}\frac{w}{\ln\left(\frac{1 + E}{1 - E}\right)} = \frac{w}{2\ln(3)},
\end{equation}
which is shown as the bottom diagonal line in Fig.~\ref{fig:paramspace}.
Additionally, we notice that this system has an infinite horizon for all system configurations, for which the particle can move between the potential wells.
One example of these horizons is shown as the dotted line in Fig.~\ref{fig:2dpot}.

As the wells get closer together or the potential becomes softer, the maxima of the potential decrease, eventually becoming lower than the total energy of the particle ($E = 1/2$). This scenario cannot be realized in a repulsive bump system with Fermi-type scatterers by varying only the parameters $w$ and $\sigma$. In contrast, the potential well configuration considered here enables access to a broader range of dynamical behaviors through parameter tuning alone.
We can obtain an approximate condition for this limit by considering the effects of three potential wells arranged in a triangular configuration, resulting in
\begin{equation}
    \sigma \gtrsim \frac{2\sqrt{3} - 3 + \sqrt{3}w}{3\ln(5)}.
\end{equation}
This limit is denoted with the upper line in Fig.~\ref{fig:paramspace}.
This Figure also includes two dashed lines and one dotted line, representing areas of further analysis discussed in Sec.~\ref{sec:results}A.
\revision{These areas were selected as representative slices through various parameter space structures further discussed in Sec.~\ref{sec:results}B.}

\subsection{Approximations for $D$}

Several analytical approaches have been developed to estimate the diffusion coefficient $D$ in the conventional Lorentz gas, as detailed in Refs.~\cite{Klages_book,machta_zwanzig,klages2000,Dettmann_2014,Klages2002,GilSan09,GNS11,CGLS14,CGLS15}. Many of these methods are based on random walk models coupled with diffusive assumptions, expressing $D$ in terms of the scatterer density. In this work, we focus on two specific approaches: the standard MZ approximation and an extended version that incorporates hopping paths of length $n$, thereby capturing correlations between successive transitions between traps.

Although the lattice is now inverted, the underlying logic of the MZ approximation~\cite{machta_zwanzig} is applicable by associating diffusion with random hopping between neighboring traps. In this context, a trap refers to a single potential well, which has six possible exit directions corresponding to the surrounding lattice geometry. By the phase space argument, the mean residence time within a trap is approximately
\begin{equation}
    \tau \approx \frac{\Omega}{\omega},
\end{equation}
where~$\Omega$ is the total allowed phase space volume within the trap and~$\omega$ is the exiting phase space flux from the trap.
Integrating over both velocity and position spaces gives
\begin{align}
    \Omega &= 2\pi \int_{V(\vect{r}) \leq E} v(\vect{r}) \dd{\vect{r}} = 2\pi A_{\text{trap}} \mean{v(\vect{r})}_{\text{trap}}, 
\end{align}
where~$v(\vect{r}) = \sqrt{2\left(E - V(\vect{r})\right)}$ is the magnitude of the velocity obtained from the conservation of energy $E$, $A_{\text{trap}}$ is the area of the trap, and $\mean{v(\vect{r})}_{\text{trap}}$ is the average magnitude of the velocity within the trap.
As we defined the trap to be a single potential well, we now have six exits, and through symmetry we obtain
\begin{align}
    \omega &= 6 \int_{V(y) \leq E} \int_{-\frac{1}{2}\pi}^{\frac{1}{2}\pi} v(y)^2 \cos\theta \dd\theta \dd y \\
    &= 12 l_{\text{exit}} \mean{v(\vect{r})^2}_{\text{exit}} \text{,}
\end{align}
where $V(y)$ and $v(y)$ are evaluated at $\vect{r}=(r_0, y)$, $l_{\text{exit}}$ is the length of the exit, and $\mean{v(\vect{r})^2}_{\text{exit}}$ is the mean squared magnitude of the velocity within the exit. Note that the integrand represents the velocity-space density multiplied by the dot product of the velocity vector and the outward-pointing unit normal of the exit. This yields the outward phase-space flux at position $x$ in direction $\theta$, and the total flux upon integration.

Finally, we substitute our values into the MZ diffusion coefficient approximation, and obtain
\begin{equation}
    D_\mathrm{MZ,soft} = \frac{l^2}{4\tau} = \frac{3\left(2r_0 + w\right)^2 l_{\text{exit}} \mean{v(\vect{r})^2}_{\text{exit}}}{2\pi A_{\text{trap}} \mean{v(\vect{r})}_{\text{trap}}},
\end{equation}
where the distance between traps $l$ is equal to the lattice spacing.
No analytical solutions for~$A_\mathrm{trap}$, $l_\mathrm{exit}$ or the averages can be obtained, but they can be computed numerically.

In the following, we adopt the hopping model introduced in Ref.~\cite{toivonen2025} to derive a more accurate expression for the diffusion coefficient.
In this approach, trap-hopping sequences of length $n$ are considered, and the diffusion coefficient $D$ is computed based on the lengths and durations of these paths, weighted by their respective probabilities. Each hop is defined as an exit from a trap in a specific absolute direction.
Our trap has six exits, and we define~$\mathcal{H} = \{\nwarrow, \nearrow, \rightarrow, \searrow, \swarrow, \leftarrow\}$ as the alphabet of possible hopping directions.
The direction of the~$i$th hop is denoted with~$\Delta_i\in\mathcal{H}$, and we can determine the displacement vector of a path~$\Delta_1\Delta_2\ldots\Delta_n$ as
\begin{equation}
    \boldsymbol{R}(\Delta_1\ldots\Delta_n) = \sum_{i=1}^n \boldsymbol{\rho}_{\Delta_i},
\end{equation}
where
\begin{align}
     \boldsymbol{\rho}_{\nwarrow}    &= \begin{bmatrix}  -1/2 \\  \sqrt{3}/2 \end{bmatrix},
    &\boldsymbol{\rho}_{\nearrow}    &= \begin{bmatrix}   1/2 \\  \sqrt{3}/2 \end{bmatrix},
    &\boldsymbol{\rho}_{\rightarrow} &= \begin{bmatrix}     1 \\  0          \end{bmatrix},\nonumber\\
     \boldsymbol{\rho}_{\searrow}    &= \begin{bmatrix}   1/2 \\ -\sqrt{3}/2 \end{bmatrix},
    &\boldsymbol{\rho}_{\swarrow}    &= \begin{bmatrix}  -1/2 \\ -\sqrt{3}/2 \end{bmatrix},
    &\boldsymbol{\rho}_{\leftarrow}  &= \begin{bmatrix}    -1 \\  0          \end{bmatrix}
\end{align}
are the unit hopping vectors between adjacent traps.
We denote the squared displacement of a hopping path as $\boldsymbol{R}^2(\Delta_1\ldots\Delta_n)$, where the square implies the vector dot product. The time-dependent diffusion coefficient is then computed by summing over all possible $n$-step hopping paths, each weighted by its probability. Since the residence time varies between paths, we account for path-dependent durations explicitly. This formulation implicitly assumes that the MSD of any individual $n$-hop path scales linearly with time over the duration of those $n$ hops. Finally, we obtain
\begin{equation}\label{eq:hopping}
    D_\mathrm{MZ}^{n\mathrm{-hop}} = \frac{l^2}{4} \sum_{\substack{\text{Permutations}\\\Delta_1 \cdots \Delta_n}} \frac{ p(\Delta_1 \cdots \Delta_n) \boldsymbol{R}^2(\Delta_1 \cdots \Delta_n) }{\tau(\Delta_1 \cdots \Delta_n)},
\end{equation}
where~$p(\Delta_1 \cdots \Delta_n)$ is the probability of a certain path and~$\tau(\Delta_1 \cdots \Delta_n)$ is the time taken by the path.
The sum runs over all unique hopping paths of length~$n$, and the resulting approximation for the diffusion coefficient is denoted by~$D_\mathrm{MZ}^{n\mathrm{-hop}}$. Since no analytical expressions are available, the probabilities and residence times are estimated from simulation data.

However, this hopping model excludes trajectories that remain confined within a single potential well -- including some of the LPOs -- as they do not contribute to hopping sequences.
To correct for their effect, we can derive a correction term starting from the MSD.
Denoting the number of non-confined trajectories as~$N$ and the number of COs as~$n$, and assuming that the contribution of the COs to the MSD is zero, the MSD can be written as
\begin{equation}\label{eq:msd_normal}
    \mathrm{MSD} = \frac{\left[\sum^N(\boldsymbol{r}(t) - \boldsymbol{r}(0))^2\right] + n \times 0}{N + n},
\end{equation}
where the sum runs over all non-confined trajectories.
We note that the MSD calculated in Eq.~\eqref{eq:msd_normal} corresponds to the MSD of the numerical diffusion coefficient calculations.
However, if we only consider the non-confined trajectories in the MSD calculation, we have
\begin{equation}\label{eq:msd_no_co}
    \mathrm{MSD}_\mathrm{non-CO} = \frac{\sum^N(\boldsymbol{r}(t) - \boldsymbol{r}(0))^2}{N}.
\end{equation}
This is the MSD that is implicitly utilized while computing~$D_\mathrm{MZ}^{n-\mathrm{hop}}$.

If we now divide Eqs.~\eqref{eq:msd_normal} and~\eqref{eq:msd_no_co}, we can obtain a correction term
\begin{align}
    \frac{\mathrm{MSD}}{\mathrm{MSD}_\mathrm{non-CO}} &= \frac{N}{N + n} = 1 - \frac{n}{N + n} = 1 - \rho_\mathrm{CO},
\end{align}
where we define~$\rho_\mathrm{CO}$ as the fraction of confined orbits, obtained through simulations.

By applying this correction term, we can define a CO-corrected approximation as
\begin{equation}\label{eq:hoppingcorr}
    D_\mathrm{MZ,CO}^{n\mathrm{-hop}} = (1 - \rho_\mathrm{CO})D_\mathrm{MZ}^{n\mathrm{-hop}}.
\end{equation}
Furthermore, we can remove the effect of COs from the numerical diffusion coefficient by dividing it by our correction term, which we denote with~$D_\mathrm{CO}$.

\section{Results}\label{sec:results}

\subsection{Diffusion coefficient}

\begin{figure}[htb]
    \centering
    \includegraphics{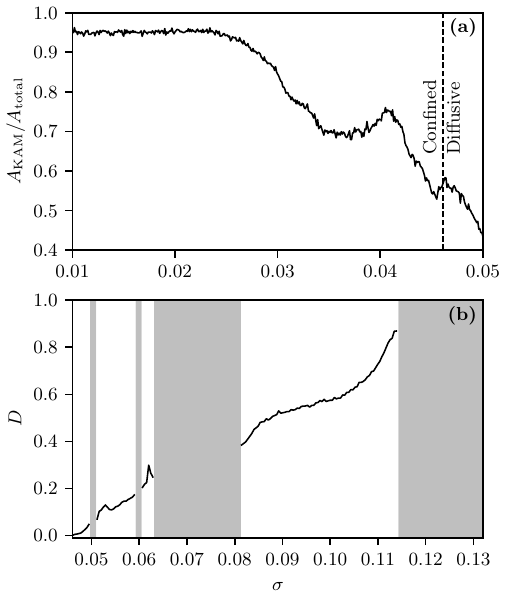}
    \caption{
        (a) Fractional area of regular motion in the Poincaré surface of section (PSOS) as a function of~$\sigma$ at~$w = 0.1013$. Chaotic dynamics is detected with the 0-1 test for chaos~\cite{gottwald2004,gottwald2009a,gottwald2009b}. The system exhibits chaos even before the particles can escape the potential wells. Below~$\sigma\approx 0.025$, the system is fully regular, and the discrepancies are caused by the false positives of the algorithm. The analyzed parameter space is marked in Fig.~\ref{fig:paramspace} as a white dotted vertical line.
        (b) Diffusion coefficient as a function of $\sigma$ at~$w = 0.1013$. In the grey areas, the diffusion coefficient is not defined due to the presence of quasiballistic orbits. The analyzed parameter space is marked in Fig.~\ref{fig:paramspace} as a black dashed vertical line.
    }
    \label{fig:dcoef_s}
\end{figure}

We examine the diffusion coefficient as a function of both $w$ and $\sigma$. In Fig.~\ref{fig:dcoef_s}(a), we focus on the system’s chaotic behavior prior to reaching the diffusive regime, which is marked by the dashed vertical line. For this analysis, we used an ensemble of $N = 3000$ trajectories with a maximum simulation time of $T = 5000$. The figure shows, as a function of~$\sigma$ at~$w = 0.1013$, the relative fraction of the accessible area in the Poincaré surface of section (PSOS) that is occupied by regular trajectories, as identified using the 0–1 test for chaos~\cite{gottwald2004,gottwald2009a,gottwald2009b}.
We have defined our PSOS as the main diagonal of a parallelepiped unit cell as shown in Fig.~\ref{fig:2dpot}.

Our results indicate that chaotic behavior begins to emerge well before the onset of diffusion, at around~\hbox{$\sigma \approx 0.025$}. Below this threshold, the PSOS appears entirely regular, which we confirmed through visual inspection. Minor discrepancies in the Figure are attributed to false positives produced by the 0–1 chaos detection algorithm~\cite{marszalek2019}.

\begin{figure*}[htb]
    \centering
    \includegraphics{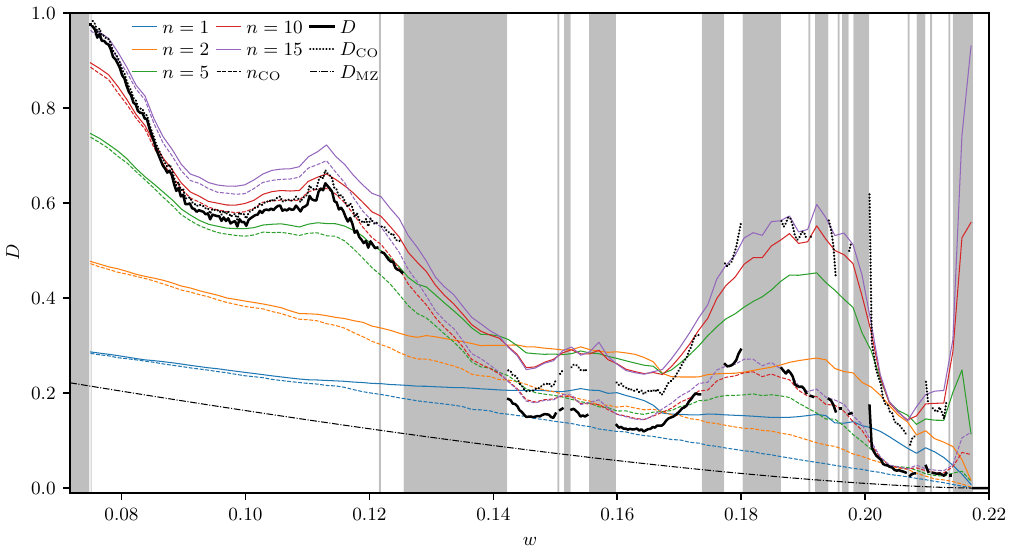}
    \caption{
        Diffusion coefficient~$D$ as a function of gap width $w$ at $\sigma = 0.0989$ (thick, black, uniform line).
        The Machta-Zwanzig approximation is presented as the dash-dotted line, and the hopping model approximations [Eq.~\eqref{eq:hopping}] are presented as colored lines.
        The dashed lines represent CO-corrected values [Eq.~\eqref{eq:hoppingcorr}] and the dotted line the corrected diffusion coefficient.
        The grey stripes indicate parameter areas with quasiballistic trajectories.
        The analyzed parameter space is included in Fig.~\ref{fig:paramspace} as a black dashed horizontal line.
    }
    \label{fig:dcoef_w}
\end{figure*}

In Fig.~\ref{fig:dcoef_s}(b) we present the diffusion coefficient as a function of system softness~$\sigma$ at~$w = 0.1013$.
In these simulations,~$N=\num{40000}$ and~$T=\num{40000}$ for each value of~$\sigma$.
We note that the diffusion coefficient has a complex dependency on the system parameters. Overall, it tends to increase with growing softness, as this leads to wider and shallower gaps between potential wells, facilitating particle transport. Notably, there are broad regions in parameter space where the diffusion coefficient is undefined due to the emergence of anomalous diffusion, caused by quasiballistic trajectories that violate the assumptions of normal diffusive behavior. These areas are marked with grey in the Figure.

In Fig.~\ref{fig:dcoef_w}, the diffusion coefficient is presented as a function of $w$ for $\sigma = 0.0989$.
Once again, $D$ is a complex function of~$w$.
In the simulations,~$N=T=\num{40000}$.
Overall, as the spacing between wells increases, the diffusion coefficient tends to decline. This is caused by the narrowing gap where the particles can escape the well.
Just before~$w = 0.22$, the gap closes, preventing any further diffusion.
Furthermore, several gray regions indicate parameter ranges where $D$ is undefined due to the presence of quasiballistic trajectories that give rise to anomalous diffusion.

In Fig.~\ref{fig:dcoef_w}, we also present $D_\mathrm{MZ}$ as a dash-dotted line, and note that it underestimates the diffusion coefficient even though the onset of diffusion is captured.
Moreover, we present $D_\mathrm{MZ}^{n\mathrm{-hop}}$ for several values of~$n$ as colored lines. Increasing the number of hops generally increases the value of the approximate diffusion coefficient.
However, the approximation overshoots the actual diffusion coefficient, at least partly caused by the numerous COs in the lattice.
When applying the correction in Eq.~\eqref{eq:hoppingcorr}, presented as colored dashed lines, we see that the approximation becomes closer to the actual value of~$D$, especially between~$w = 0.17-0.22$.
We can also apply similar corrections to the computed diffusion coefficient by removing COs from the calculations. This result is shown as a dotted line in Fig.~\ref{fig:dcoef_w}. Now, the corrected diffusion coefficient is close to the approximated value, especially between~$w = 0.17-0.22$.

\revision{However, despite correcting for COs, a slight disagreement between the numerical diffusion coefficient and the approximation remains. This is likely related to the complex phase-space structure with localized orbits, which hinders precise convergence. Thus, the approximation captures only the qualitative behavior, and the exact cause of the discrepancy is left for future work.}

\subsection{Parameter space}

\begin{figure}[htb]
    \centering
    \includegraphics{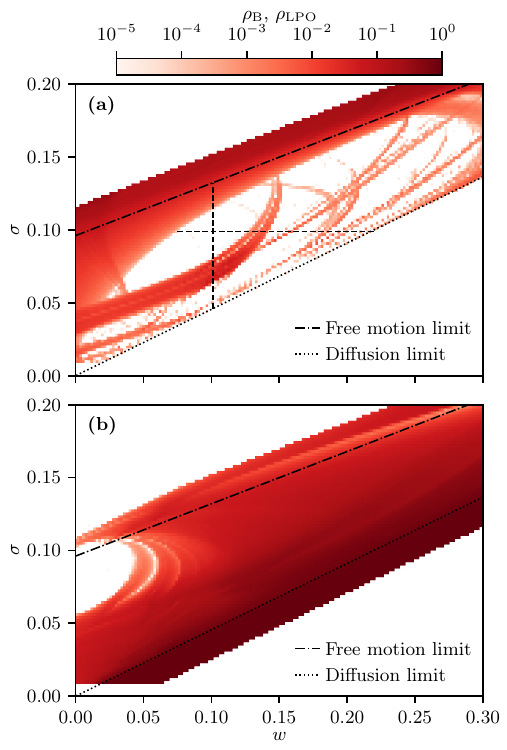}
    \caption{
        (a) Density of quasiballistic trajectories ($\rho_\mathrm{B}$) in the parameter space of the inverted triangular soft Lorentz gas, characterized by a tongue-like structure inside a lagoon without quasiballistic trajectories.
        \revision{The dashed vertical and horizontal lines refer to the parameter areas studied in Figs.~\ref{fig:dcoef_s}(b) and~\ref{fig:dcoef_w}, respectively.}
        (b) Density of localized periodic orbits ($\rho_\mathrm{LPO}$) over the parameter space. LPOs are ubiquitous over almost the whole parameter space.
        The diffusion and free motion limits correspond to the lines in Fig.~\ref{fig:paramspace}.
    }
    \label{fig:tongue_ball}
\end{figure}

First, we study the proportion of quasiballistic trajectories ($\rho_\mathrm{B}$) in Fig.~\ref{fig:tongue_ball}(a).
In these simulations,~$N=\num{100000}$ and~$T=\num{1000}$.
The dashed and dotted lines represent the free movement and diffusion thresholds, respectively, cf. Fig.~\ref{fig:paramspace}.
We did not conduct simulations in the white region at the top of the figure, as particles can move freely there, making the dynamics trivial from a diffusive perspective. Similarly, simulations were omitted below the threshold where particles are unable to escape the wells, since diffusion is entirely suppressed in that regime.

We notice that the quasiballistic trajectories of Fig.~\ref{fig:tongue_ball}(a) form \emph{similar tongue structures} compared to earlier studies~\cite{PRL,toivonen2025}.
However, we see non-ballistic motion \emph{only} in the lagoon-like area shown in the figure.
Further simulations with larger values of~$w$ revealed only more ballistic motion.
Moreover, several of the tongues are relatively thin, suggesting that even more simulations are needed to fully characterize the behavior of the system in the parameter space.
In the area where the particles cannot escape the potential well, no diffusion occurs and therefore also no quasiballistic motion is observed.

The quasiballistic orbits and LPOs in Fig.~\ref{fig:tongue_ball} have been extracted with the following scheme: First, the distance traveled by a particle was detrended with a linear fit.
Then, the standard deviation of the detrended time series was calculated.
Trajectories with a standard deviation less than one were classified as periodic. Quasiballistic and localized orbits were further distinguished based on the maximum distance traveled, using a threshold value of 25.

Next, we examine the structures formed by LPOs within the parameter space shown in Fig.~\ref{fig:tongue_ball}(b) as proportion of LPOs ($\rho_\mathrm{LPO}$).
As before, simulations were not performed in the top-left and bottom regions of the figure.
First, we observe that these orbits are widespread across the entire simulated parameter space, with no tongue-like structures. However, a small region appears to be free of LPOs, located at small values of $w$ and relatively high values of $\sigma$ (around 0.1).

\begin{figure}[htb]
    \centering
    \includegraphics{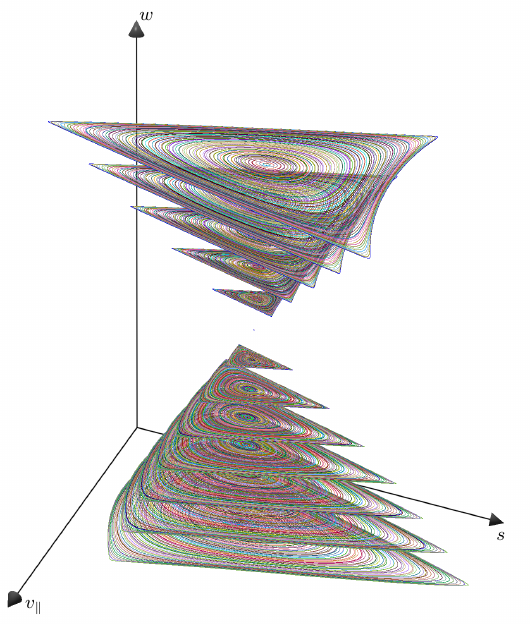}
    % \begin{tikzpicture}
    %     \node[inner sep=0pt]  at (0,0) {\includegraphics[width=.5\textwidth]{fig8.png}};
    %     \node at (-2.7, 5.5) {$w$};
    %     \node at (4, -3) {$s$};
    %     \node at (-3.6, -5.5) {$v_\parallel$};
    % \end{tikzpicture}
    \caption{
        Example of a bifurcation visible in the Poincaré surface of section as $w$ is varied, ranging from 0.0455 to 0.0543 at~$\sigma\approx 0.0763$.
        The KAM island corresponds to LPOs.
    }
    \label{fig:psos}
\end{figure}

When summing the densities of quasiballistic orbits and LPOs together, we find that regularity is present everywhere in the parameter space even though there are areas without quasiballistic orbits or LPOs.
There is a small area where the density of regular trajectories seems to vanish. This is related to a bifurcation where a KAM island consisting of LPOs is flipped as shown in Fig.~\ref{fig:psos}.
The~$s$ and~$v_\parallel$ axes refer to the distance along the diagonal of the parallelepiped at the time of crossing and the velocity component parallel to the diagonal, respectively.

\subsection{Example trajectories}

\begin{figure*}[htb]
    \centering
    \includegraphics{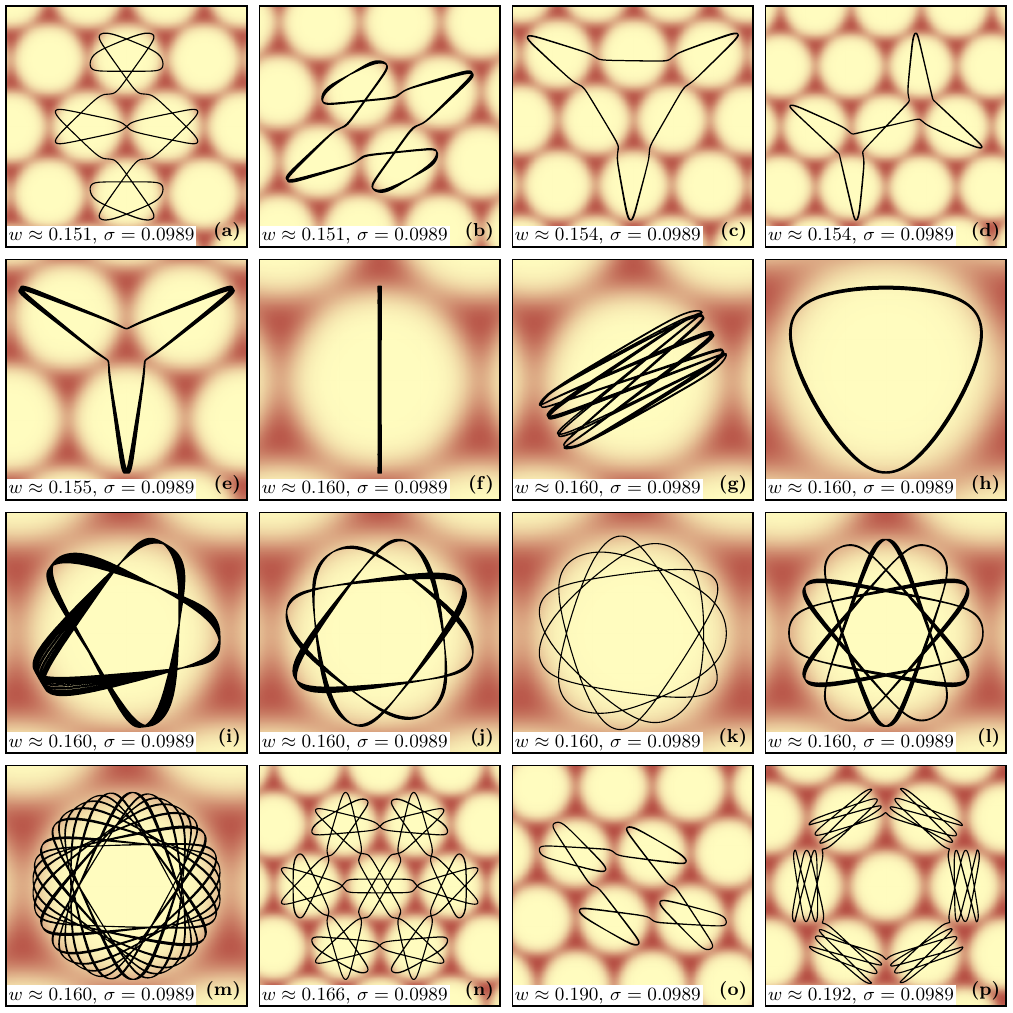}
    \caption{
        Examples of the rich selection of localized periodic orbits (LPOs) present in the inverted triangular soft Lorentz gas.
        (a-e, o) LPOs spanning a couple of potential wells.
        (f-m) LPOs confined in a single well.
        (n, p) Extremely complex LPOs traveling through 6 and 7 potential wells.
    }
    \label{fig:trajs}
\end{figure*}

In Fig.~\ref{fig:trajs}, we present some examples of LPOs present in the system.
We observe a diverse range of LPOs, from trajectories confined to a single potential well, as illustrated in Fig.~\ref{fig:trajs}(f–m), to those spanning a few neighboring wells, shown in Fig.~\ref{fig:trajs}(a–e) and (o). Additionally, we identify highly intricate orbits that traverse multiple wells across the lattice, exemplified in Fig.~\ref{fig:trajs}(n) and (p). We simulated these trajectories to a maximum time of~\num{40000}, confirming the robustness of our algorithms and the stability of the trajectories.

For the single-well LPOs, we find trajectories with different winding numbers and shapes.
First of all, we have a line-like trajectory in Fig.~\ref{fig:trajs}(f).
Furthermore, in Fig.~\ref{fig:trajs}(h) we present a triangular-ish trajectory.
Finally, in Fig.~\ref{fig:trajs}(i-m) we show trajectories with increasingly high winding numbers.

For the more complex LPOs, we present a simple three-pronged example in Fig.~\ref{fig:trajs}(e), with a larger similar example in Fig.~\ref{fig:trajs}(c).
In Fig.~\ref{fig:trajs}(d), the components of the trajectory are similar, however due to the crossing in the centermost potential well, the shape is no longer triangular.
In Fig.~\ref{fig:trajs}(a-b, o), we present some multi-well LPOs which look like precursors to Fig.~\ref{fig:trajs}(n), especially the one in Fig.~\ref{fig:trajs}(a).

Finally, for extremely long periods, we present the LPOs in Fig.~\ref{fig:trajs}(n, p).
These complex trajectories in Fig.~\ref{fig:trajs}(n) and (p) reached over 300 periods during the simulation time, confirming their periodicity and affirming the numerical stability of our propagation code.

\begin{figure}[htb]
    \centering
    \includegraphics{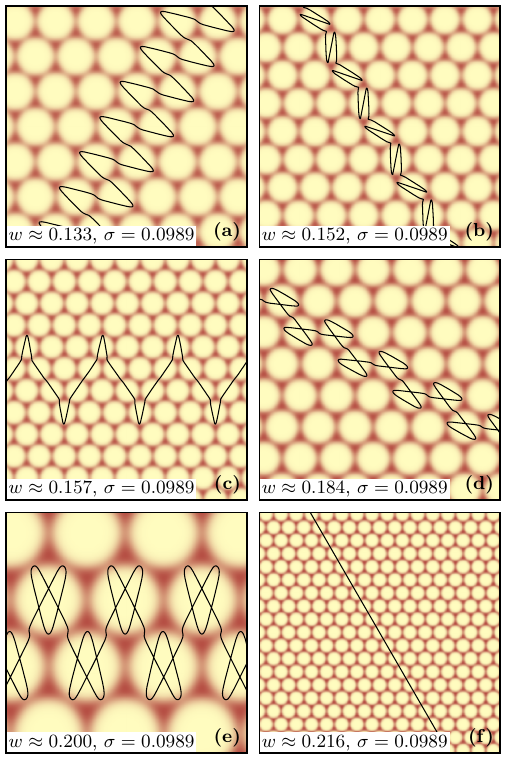}
    \caption{
        Examples of the quasiballistic trajectories found in the inverted triangular soft Lorentz gas, ranging from simple straight lines in (f) to complex intermediate shapes between (a-e).
    }
    \label{fig:ballistic}
\end{figure}

Some examples of quasiballistic orbits are presented in Fig.~\ref{fig:ballistic}.
We find trajectories ranging from simple straight lines [Fig.~\ref{fig:ballistic}(f)] to complex intermediate shapes [Fig.~\ref{fig:ballistic}(e)]

\section{Conclusions}\label{sec:conclusion}

In this work, we have investigated the diffusion properties of an inverted triangular soft Lorentz gas, combining analytical modeling with numerical simulations under systematic variation of key parameters.
Compared to systems with repulsive bumps, this inverted configuration allows for richer control over particle transport, as free motion can be achieved through system parameter tuning without changing the energy of the particle.

Building on our earlier hopping model for approximating the diffusion coefficient, we introduced a refined version that accounts for the influence of confined orbits (COs). This correction is essential, as COs are abundant across nearly all regions of the parameter space in the inverted system. The improved model significantly reduces the discrepancy between predictions and numerical results for the diffusion coefficient.

Our results reveal that anomalous diffusion, particularly associated with localized periodic orbits, is widespread and exhibits a rich spectrum of dynamical behavior. We observed trajectory types ranging from those confined within a single potential well to highly complex orbits traversing multiple wells. In addition, quasiballistic trajectories form intricate, tongue-like structures in parameter space, signaling complex transport regimes that challenge standard diffusion theory.

Importantly, we found that the system exhibits chaotic dynamics even before the onset of diffusion, driven by the coupling between adjacent potential wells. This suggests that signatures of chaos may emerge earlier than expected in systems with soft, smooth potentials -- as suggested in Ref.~\cite{gonzales_arxiv}.

From a broader perspective, this model system is timely and relevant in light of recent advances in fabricating two-dimensional materials with engineered potential landscapes, such as those designed for nanoparticle trapping or electron transport in artificial graphene. Our work thus contributes to the growing theoretical foundation needed to understand diffusion in such tailored soft-matter and quantum systems. By refining and applying our hopping-based framework, we provide new tools for characterizing complex transport phenomena in smooth, structured potentials, hopefully opening avenues for both classical and quantum applications.

% If you have acknowledgments, this puts in the proper section head.
\begin{acknowledgments}
This work was funded by the Research Council of Finland, ManyBody2D Project (No. 349956).
The authors wish to acknowledge CSC – IT Center for Science, Finland, for computational resources.
The authors thank Balázs Paszkál Halmos for performing numerical simulations shown in Fig.~\ref{fig:dcoef_s}(a).
\end{acknowledgments}

% Create the reference section using BibTeX:
\bibliography{references}

\end{document}